\documentstyle[12pt]{article}
    \newcommand{\be}{\begin{equation}}
    \newcommand{\ee}{\end{equation}}
    \def\n{\noindent}
    \catcode `\@=11
    \catcode `\@=12

    \title{\bf\huge Electromagnetic duality in general relativity}
    
    \author{Naresh Dadhich\thanks{E-mail : nkd@iucaa.ernet.in} \\
    {\sl Inter-University Centre for Astronomy \& Astrophysics,}\\
    {\sl Post Bag 4, Ganeshkhind, Pune - 411 007, India.} 
    } 
    \date{}
    \begin{document}
    \maketitle
    
     \begin{abstract}

    By resolving the  Riemann curvature relative to a unit timelike vector 
   into electric and magnetic parts, we consider duality relations 
analogous to the electromagnetic theory. It turns out that the duality 
symmetry of the Einstein action implies the Einstein vacuum equation 
without the cosmological term. The vacuum equation is invariant 
under interchange of active and passive electric parts giving rise to the 
same vacuum solutions but the gravitational constant changes sign. 
Further by modifying the equation it is possible to construct 
interesting dual solutions to vacuum as well as to flat spacetimes.
    \end{abstract}

    \n PACS numbers : 04.20,04.60,98.80Hw

    \newpage 
    
\section{Introduction}

   In analogy with  the  electromagnetic  field, it is possible to resolve 
the gravitational field;, i.e.\ Riemann curvature tensor,  into electric and 
magnetic parts relative to  a unit  timelike  vector [1-4]. In general, 
a field is produced by its charge (source). Its manifestation when 
charge is stationary is termed electric, and magnetic when it is 
moving. The Maxwell electromagnetic field was the first example that 
brought 
forth this general feature and hence provided the terminology for other 
fields. It should be 
recognised that this is a general property of any classical field. 

 In general relativity (GR), unlike other fields, charge is also of two 
kinds. In addition to the usual charge in terms of the usual non-gravitational
matter/energy distribution, gravitational field energy 
itself also has charge. This is what makes the theory 
non-linear. Thus the electric part would also be of two kinds corresponding to 
the two kinds of charge, which we term active (non-gravitational energy) 
and passive (gravitational field energy). 
    
 The 20 components of the Riemann curvature are split into 6 each of 
active (projection of Riemann, $R_{abcd}u^bu^d$) and passive (projection 
of left and right dual) parts and 8 of magnetic (projection of left 
or right dual) part. The electric and magnetic parts are second rank 
3-space tensors orthogonal to the resolving timelike unit vector $u^a$, the 
electric parts are symmetric while the magnetic part is trace free and
is the
sum of the symmetric Weyl magnetic part  and the anti-symmetric part 
representing energy flux. 

 Clearly gravitational field has richer structure than electromagnetic 
field. It would be interesting to see what the duality realtion 
involving electric and magnetic parts implies? It turns out that the 
duality transformation, analogous to the Maxwell theory, which keeps the 
Einstein action invariant interestingly implies the Einstein vacuum 
equation [5-6]. Note that there is a basic difference between gravitational 
and 
electromagnetic fields. For the former, the Riemann curvature contains 
the entire dynamics (field equation) as it involves second order 
derivatives of the metric (potential), while for the latter fields are 
first order derivatives of the gauge potential and the dynamics (field 
equation) would follow from derivatives of the fields. Hence it is 
understandable that any manipulation of the Riemann curvature would 
always refer to dynamics of gravitational field. 

 Remarkably the gravo-electromagnetic duality symmetry of the Einstein 
action implies the Einstein vacuum equation without the cosmological 
constant (The equation with cosmological constant is characterized by 
equality of left and right dual of the Riemann curvature). This 
property is similar to the other well-known property of GR that the field 
equation implies the equation of 
motion for free particles. Now we have the symmetry of the action implying 
the equation of motion for the field. In GR, there is 
always synthesis of physical quantities, concepts and equations. 

 \n The Einstein vacuum equation, written in terms of electric and magnetic 
parts, is symmetric in active and passive electric parts. We consider 
another duality relation, which we call gravo-electric duality, an 
interchange of active and passive electric parts which keeps the vacuum 
equation invariant. Under this transformation it turns out that 
the Ricci and the Einstein tensors are dual to each other. That is, the 
non-vacuum equation will in general distinguish between active and passive 
parts. There could occur solutions that are dual to each other [7]. In 
particular it follows that perfect fluid spacetimes with the equations of 
state $\rho - 3p = 0$ and $\rho + p = 0$ ($\Lambda \rightarrow 
-\Lambda$) are self dual while the stiff fluid is dual to dust.

 \n Under the gravo-electric duality though, the vacuum equation remains 
invariant yielding the same vacuum solutions, but the gravitational constant 
$G$ will change sign. This is because the scalar curvature $R$ changes 
sign and now to keep action invariant $G$ must change sign. In obtaining 
vacuum solutions for isolated bodies, there always remains one equation 
free which is implied by the others. If we now tamper with this equation the 
vacuum solution will remain undisturbed but this would make the 
equation non-invariant under the gravo-electric duality. Thus by 
modifying the vacuum equation suitably distinct solutions dual to 
the well-known black hole solutions could be obtained.
   
 \n In sec. 2, we shall give the electromagnetic decomposition of the 
Riemann curvature, followed by the duality symmetry of the 
Einstein action implying the vacuum equation in Sec. 3. In 
Sec. 4 we shall discuss the duality transformation that keeps the vacuum 
equation invariant and its implication for the black hole spacetimes. By 
modifying the vacuum equation it is possible to find interesting solutions 
dual to 
the black hole spacetimes which will be discussed in Sec.5. Finally we 
conclude with discussion.

\section{Electromagnetic decomposition}

   \n We resolve the  Riemann  curvature tensor relative to a unit timelike 
vector as follows :
    
    \be
    E_{ac} = R_{abcd} u^b u^d,  \tilde E_{ac} = *R*_{abcd} u^b u^d
    \ee
    
    \be
    H_{ac} = *R_{abcd} u^b u^d = H_{(ac)} - H_{[ac]}, H_{ca} = 
R*_{abcd}u^bu^d = \tilde H_{ac} \\ 
    \ee
    
    \n where
    
    \be
    H_{(ac)} = *C_{abcd} u^b u^d
    \ee
    
    \be
    H_{[ac]} = \frac{1}{2} \eta_{abce} R^e_d u^b u^d.
    \ee
    
    \n Here $C_{abcd}$  is  the  Weyl conformal  curvature, $\eta_{abcd}$
    is  the  4-dimensional volume element.  Note that the magnetic part 
is the projection of left ($H_{ac}$) or right ($H_{ca}$) dual and hence 
either one of them can be taken. We shall therefore drop $\tilde H_{ac}$ 
from further discussion. We have $E_{ab} = E_{ba}, {\tilde E}_{ab}
    = {\tilde E}_{ba}, (E_{ab}, {\tilde E}_{ab}, H_{ab})
    u^b = 0,~ H= H^a_a = 0$ and $u^a u_a = 1$. 
    The Ricci tensor could then be written as
    
    \be
    R_{ab} = E_{ab} + {\tilde E}_{ab} + (E + {\tilde E}) u_a u_b -
    {\tilde E } g_{ab} + \frac{1}{2} H^{mn} u^c (\eta_{acmn} u_b + 
\eta_{bcmn} u_a)
    \ee
    
    \n where $E = E^a_a$        and  $\tilde E = \tilde E^a_a$. 
    It may be noted that in view of $G_{ab} = - T_{ab}$, $E = (\tilde E + 
\frac{1}{2} T)/2$            defines 
    the  gravitational  charge density  while ${\tilde E}= - T_{ab}
    u^a u^b$            defines  the 
    energy density relative to the unit timelike vector $u^a$.   \\

 \n In terms of electromagnetic parts, the vacuum equation $R_{ab} = 0$ 
would thus read for any unit timelike resolving  vector as

\be
H_{[ab]} = 0, ~E ~or~ \tilde E = 0, ~E_{ab} + \tilde E_{ab} = 0. 
\ee

 \n It is symmetric in active and passive electric parts.

  \section{Gravo-electromagnetic duality}

 \n In electromagnetics the duality transformation $\bf E\rightarrow\bf 
H$, $\bf H\rightarrow -\bf E$ keeps the Maxwell action, which goes as $E^2 
- H^2$, and the source free Maxwell equation invariant. Note 
that this transformation would essentially lead to vacuous field, and 
hence there cannot occur a solution obeying it. This is however a symmetry 
of the action.

 \n In GR, on the other hand, all vacuum solutions would respect the 
analogous duality transformation because the vacuum equation is implied by 
the transformation. Analogously we consider
\be
E_{ab} \rightarrow H_{ab} 
\ee

\be 
H_{ab} \rightarrow - {\tilde E}_{ab} 
\ee

\n which would imply

\be
{\tilde E}_{ab} \rightarrow - E_{ab}.
\ee

 \n The first of the above relations implies that $E = 0$ because $H = 0$ 
always and $H_{[ab]} = 0$ because $E_{ab}$ and $\tilde E_{ab}$ are 
symmetric. These combined with the third relation are the Einstein vacuum 
equation (6). The above duality transformation thus implies the vacuum 
equation and it is a symmetry of the Einstein action as the 
scalar curvature $R$ remains invariant [5-6]. We thus have a remarkable 
result:

 {\it The above duality transformation is a symmetry of the Einstein 
action and implies the vacuum equation without the cosmological constant.}

 The corresponding result in the Maxwell theory is that the duality 
transformation is only the 
symmetry of the action and of the source free field equation but 
it does not imply the field equation. In GR we must recognise the fact that 
there is a richer structure through two kinds of electric parts and 
breakup of magnetic part into the symmetric Weyl free-field part and the 
antisymmetric energy flux part. More importantly these quantities are one 
order higher in differentiation, as they involve second derivatives of the 
metric. Hence gravitational electromagnetic parts, unlike the Maxwell case, 
contain dynamics of the field whereas in the Maxwell case 
dynamics emerges only on one more differentiation. This is the basic and 
crucial difference between  the two fields. Thus if the duality 
symmetry of the action were to imply an equation, it could only be  the 
eqution of motion (field equation) for gravitational field.  

 Note that the cosmological constant cannot appear in the vacuum equation 
as it is not sustainable by the duality transformation. It could however 
always come in as matter with its well-known specific equation of state, 
$\rho + p = 0$. It could however be characterised [1-2] by the 
following geometric condition,  
\be
{\bf*R} = {\bf R*} 
\ee
\n where ${\bf*R}$ and ${\bf R*}$ denote respectively the left and right dual 
of the Riemann curvature. In view of (2), it would imply 
\be
H_{[ab]} = 0,
\ee
\n and because $ ** = -1$ ($- {\bf R} = {\bf *R*}$),

\be
E_{ab} + \tilde E_{ab} = 0.
\ee

\n This obviously implies from (5) $R_{ab} = \Lambda g_{ab}$ with $E = 
\Lambda$. As a matter of fact we can make the following general statement:

{\it The necessary and sufficient condition for $R_{ab} = \Lambda g_{ab}$ 
is that ${\bf*R} = {\bf R*}$.}

 The sufficiency has been shown above. For the necessary condition, 
substituting eqn.(5) into $R_{ab} = \Lambda g_{ab}$, the eqns. (11) 
and (12) immediately follow. Further the eqn. (12) means
\be
({\bf R} + {\bf *R*}).u.u = 0.\\
\ee
\n Since this is to be true for any arbitrary unit timelike vector, then

\be
{\bf R} = - {\bf *R*} 
\ee

\n which would imply
\be
{\bf *R} = {\bf R*}.
\ee

\n So is proved the necessary condition. \\

 In addition to eqn. (10) if scalar curvature $R$ vanishes, then it is 
vacuum (because $R = 0$ implies $E = \tilde E$ then in view of eqn. (12),
$E = \Lambda = \tilde E = 0$). The vacuum is thus characterized by 
${\bf*R} ={\bf R*}$ and $R = 0$.

   \section{Duality transformation and vacuum}

 For the ready reference, we recall the vacuum equation (6)

    \be
    E ~ or ~ {\tilde E} = 0,~ H_{[ab]} = 0 = E_{ab} + {\tilde E}_{ab}
    \ee
    
  \n which is symmetric in $E_{ab}$ and ${\tilde E}_{ab}$.\\

 One may next ask, what keeps the vacuum equation invariant? Clearly the 
above equation is symmetric between active and passive electric parts. 
Thus in the second avatar of duality, which is termed as 
the gravo-electric duality, we define the duality transformation as 
   
     \be
     E_{ab} \longleftrightarrow {\tilde E}_{ab}, ~H_{ab} \longrightarrow - 
H_{ab}.
     \ee
   
   \n Thus the vacuum equation (6) is invariant under the duality 
   transformation (17). The vacuum equation is neutral about the 
transformation of $H_{ab}$. As shown below, it turns out that the Weyl 
electric part changes sign and hence so should the magnetic part. From eqn. 
(1) it is clear that the duality 
   transformation would map the Ricci tensor 
   into the Einstein tensor and vice-versa. This is because the contraction of 
   Riemann is Ricci while that of its double dual is Einstein. Note also that 
it maps $R$ to $- R$ because $R = -2(E - \tilde E)$.  \\   
   
 \n Even though the vacuum equation is a gravo-electric invariant, 
which would mean the vacuum solutions would also be invariant, the 
constants 
of integration may change sign. This is what really happens because the 
electric part of the Weyl curvature reads as

\be
2E_{ab}(W) = E_{ab}(TF) - \tilde E_{ab}(TF)
\ee

\n where TF stands for trace free part; i.e.

\be
E_{ab}(TF) = E_{ab} - \frac{1}{3}E h_{ab}
\ee

\n and
\be
h_{ab} = g_{ab} - u_au_b.
\ee

 \n Clearly under the duality transformation (17), the Weyl electric part 
and scalar curvature $R$ change sign. If this were also to be a
symmetry of the Einstein action, the gravitational constant $G$ must also 
change sign. Of course any symmetry of the equation derived from the 
action must also be a symmetry of the action itself. That means 
gravo-electric duality implies $G \rightarrow - G$; i.e.\ gravity changing 
its sense! This appears rather strange.
 
 \n A keener look into what produces  active and passive parts does 
illuminate the 
situation. $E_{ab}$ is produced by non-gravitational energy distribution 
while the source for $\tilde E_{ab}$ is gravitational field energy (we 
shall demonstrate this with an example below). The former is always 
positive as proved by the positive energy theorems while the latter is 
always negative for an attractive field. Now under duality we 
interchange active and passive parts which would amount to interchange of 
the two kinds of energy distributions having inherently opposite signs. 
This is why $G$ must change sign.

\n Further the vacuum equation essentially states that the contributions 
of the two kinds of charge (non-gravitational and gravitational) are
on an
equal footing and vacuum is characterized by vanishing of their sum. 
In GR, in contrast to the Newtonian theory, absence of non-gravitational 
energy distribution alone cannot define vacuum, because of the presence of 
gravitational field energy which can never be removed. Hence it has to be 
incorporated with due recognition of its opposite sign. We shall now 
demonstrate through the well-known case of the Schwarzschild particle 
that the field energy "curves" space while ordinary matter "pulls" [8].
  
   \n Consider the spherically symmetric metric,
    
    \be
    ds^2 = c^2(r,t) dt^2 - a^2(r,t) dr^2 - r^2 (d \theta^2 + \sin^2 \theta
    d \varphi^2).
    \ee

 \n It can be easily seen that for this metric $R_{01} = 0, R^0_0 = 
R^1_1$ lead to $c^2 = a^{-2} = 1 + 2\phi(r)$, and then $R^0_0 = - 
\bigtriangledown^2 \phi$. Thus we again solve the good old Laplace equation 
rather than contribution of field energy on the right. GR is however  
supposed to 
incorporate the contribution of the field energy. What really 
happens is that thecontribution of 
the field energy is accounted for by the curvature of space leaving the 
Laplace equation unaltered. This could be readily seen by setting $a = 1$ 
and then $R^0_0 = 0$ would have the field energy contribution on the 
right [8]. When $a \neq 1$, $R^0_0 = R^1_1$ washes out the field energy 
term on the right and gives $ac = 1$. This is how gravitational field 
energy ``curves" space. It is the space curvature which is represented by 
the passive electric part $\tilde E_{ab}$. The active part is due to 
space-time curvature which is anchored onto non-gravitational energy 
distribution. It is well-known that the Newtonian potential sitting in 
$g_{00}$ leads to acceleration as gradient of the potential in the 
geodesic equation.  

 \n Thus the interchange of active and passive part under duality would 
mean interchange of their sources, non-gravitational and gravitational 
energy, which have opposite sign. Since they have opposite sign, gravity 
must change its sense and hence $G \rightarrow - G$.

 \n It can be further verified that all the vacuum black hole solutions 
obey the duality transformation (17) with $G \rightarrow - G$ (for 
Riemann components see for instance [9]). Of course the charged black 
hole solution does not obey the transformation implying non-existence of 
the dual solution. On the other hand de Sitter spacetime is dual to anti de 
Sitter with $\Lambda \rightarrow - \Lambda$. 

 \n The NUT solution could be interpreted as the field 
of gravito-magnetic monopole [10]. By looking at its electric and magnetic 
parts [11], one observes that the difference between them goes to zero 
as $M \rightarrow l, l \rightarrow -M$ where $M$ and $l$ are mass and NUT 
parameters. Since it is well-known that there cannot exist real self-dual 
(${\bf R} = {\bf *R}$) solutions in GR, the NUT solution could be 
considered nearest to it as the difference between electric and magnetic 
parts vanishes for the appropriate transformation of the source 
parameters. 
 
 \n The duality transformation (17) could be cast in the continuous form 
as follows:
\be
E_a^b \longrightarrow E_a^b cos \theta + \tilde E_a^b sin \theta, \\
\tilde E_a^b \longrightarrow E_a^b sin\theta + \tilde E_a^b cos \theta \\
\ee
\n with
\be
\rho \longrightarrow \rho cos\theta +\frac{1}{2}(\rho + 3p) sin\theta, \\
p \longrightarrow p cos\theta + \frac{1}{2} (\rho - p) sin\theta. \\
\ee

 \n The discreete transformation (17) corresponds to $\theta = \pi/2$.

 \section{Solutions dual to black hole/flat spacetimes}

  \n Next the question arises, can we obtain a dual to a 
    vacuum solution? The vacuum equation is 
    symmetric in active and passive parts and hence invariant under 
    the duality transformation (17). However it turns out that in obtaining 
the well-known black hole solutions not all of the vacuum equations are 
used. In particular, for the Schwarzschild solution the equation $R_{00} = 
0$ in the standard curvature coordinates is implied by the rest of the 
equations. If we tamper with this equation, the Schwarzschild solution 
would remain undisturbed for the rest of the set will determine it 
completely. However this modification, which does not 
affect the vacuum solution, breaks the symmetry between active and 
passive electric parts leading to non-invariance of the modified equation 
under the duality transformation. This would lead to  distinct dual 
solutions. We shall demonstrate this by obtaining a dual solution to the 
Schwarzschild solution by modifying the vacuum equation appropriately.
        
    \n For the metric (21) the natural choice for the resolving vector is of 
   course the hypersurface orthogonal unit vector, pointing along the 
$t$-line. 
 From   eqn. (6), $H_{[ab]} = 0$ and $E^2_2 + {\tilde E}^2_2 = 0$ lead 
   to $ac = 1$ (for this, no boundary condition of asymptotic flatness need 
   be used [8]). Now ${\tilde E} = 0$ gives $a = (1-2M/r)^{-1/2}$, which 
determines the 
   Schwarzschild solution completely. Note that we did not need to use the 
remaining 
   equation $E^1_1 + {\tilde E}^1_1 = 0$, it is  hence free and is
   implied by the rest. Without affecting the Schwarzschild solution, we 
   can introduce some distribution in the 1-direction.      
   
   \n We hence write the alternate equation as
    
    \be
    H_{[ab]} = 0 = {\tilde E},~ E_{ab} + {\tilde E}_{ab}
    = \lambda w_a w_b
    \ee
    
    \n where $\lambda$ is a scalar and $w_a$ is a spacelike unit vector 
along the direction of $4$-acceleration. It is clear that it will also 
   admit the Schwarzschild solution as the general solution, and 
determine 
   $\lambda = 0$. That is, for spherical symmetry the above alternate 
   equation also characterizes vacuum, because the Schwarzschild solution is 
   unique. \\
   
    \n  Let us now employ the duality  transformation (17) to the above  
   equation (22) to write  
    
    \be
    H_{[ab]} = 0 = E,~ E_{ab} + {\tilde E}_{ab} = \lambda w_a w_b.
    \ee

    \n Its general solution for the metric (21) is given by 
    
    \be
    c = a^{-1} = (1 - 2k - \frac{2M}{r})^{1/2}.
    \ee

    \n This is the Barriola-Vilenkin solution [12] for the  Schwarzschild 
    particle with global monopole charge parameter, $\sqrt {2k}$. Again we 
shall 
      have $ac = 1$ and $E=0$ will then yield $c = (1-2k - 2M/r)^{1/2}$ 
      and $\lambda = 2k/r^2$. This  has  non-zero stresses given by
    
    \be
    T^0_0 = T^1_1 = \frac{2k}{r^2}.
    \ee

    \n A global  monopole is supposed to be produced by spontaneous 
breaking of the global symmetry $O(3)$ into $U(1)$ in a phase transition in 
the early Universe. It   is    described    by    a    triplet    scalar,
     $\psi^a (r) = \eta f(r) x^a/r, x^a x^a = r^2$,
    which  through  the usual Lagrangian  generates an energy-momentum 
    distribution at large distance from the core precisely of the  form 
    given  above  in (25) [12].  Like the Schwarzschild solution the 
   monopole solution (24) is also the unique solution of eqn.(23). \\ 
 
    \n If we translate eqns. (22) and (23) in terms of the familiar Ricci 
 components, they would read as
 
    \be
    R^0_0 = R^1_1 = {\lambda},   R^2_2 = 0 = R_{01}
    \ee
 
    \n and
 
    \be
    R^0_0 = R^1_1 = 0 = R_{01},  R^2_2 = {\lambda}.
     \ee
 
     \n In either case we shall have $ac = 1$ and $c^2 = f(r) 
      =1 + 2 \phi$, say, and 

     \be
     R^0_0 = - \bigtriangledown^2 \phi 
     \ee

     \be
     R^2_2 = - \frac {2}{r^2} (r \phi)^{\prime}. 
     \ee 

    \n Now (26) integrates to give ${\phi} = - M/r$ and 
 ${\lambda} = 0$, which is the Schwarzschild solution while (27) will 
 give the dual solution with ${\phi} = - k - M/r$ and ${\lambda} = 
2k/{r}^2$, the 
 Schwarzschild solution with global monopole charge. Thus the global
 monopole owes
 its existence to the constant $k$ appearing in the solution of the 
 usual Laplace equation. It defines a 
 pure gauge for the Newtonian theory, which could be chosen freely, 
 while the Einstein vacuum equation determines it to be zero. For the 
 dual-vacuum equation (23), it is free like the Newtonian case but it 
 produces non-zero curvature and hence would represent non-trivial 
 physical and dynamical effects (see $R^2_2 = - 2k/{r}^2 \neq 0$ 
 unless $k = 0$). This is the crucial difference between the 
 Newtonian theory and GR in relation to this problem, that the latter 
 determines the relativistic potential ${\phi}$ absolutely, vanishing 
 only at infinity. The freedom of choosing zero of the potential is 
restored in the dual-vacuum equation, 
 of course at the cost of introducing stresses that represent a global 
 monopole charge. The uniform potential would hence represent a massless 
 global monopole ($M = 0$ in the solution (24)), which is solely 
 supported by the passive part of electric field. As has been argued and 
shown above, it is the non-linear aspect of the field (which 
 incorporates interaction of gravitational field energy density) that 
 produces space-curvatures and consequently the passive electric part. It 
 is important to note that the relativistic potential ${\phi}$ plays the 
 dual role of the Newtonian potential as well as the non-Newtonian role 
 of producing curvature in space. The latter aspect persists even when 
 potential is constant different from zero. It is the dual-vacuum 
 equation that uncovers this aspect of the field. \\
   
     \n On the other hand, flat spacetime could also in alternative form be 
   characterized by 
    \be
    {\tilde E}_{ab} = 0 = H_{[ab]}, E_{ab} = \lambda w_a w_b
    \ee
    
    \n leading to $c=a=1$, and implying ${\lambda} = 0$ . Its dual will be 
    
    \be
    E_{ab} = 0 = H_{[ab]}, {\tilde E}_{ab} = \lambda w_a w_b
    \ee

    \n yielding the general solution,
    
    \be
    c^{\prime} = a^{\prime} = 0 \Longrightarrow c=1, a = const. = (1-2k)^{-1/2}
    \ee

    \n which is non-flat and represents a zero mass global monopole, 
    as follows from the solution (24) when $M=0$. This is also the 
 uniform relativistic potential solution. It can naturally be envisioned 
as "minimally" curved spacetime, which was first considered by the author 
[13] long back. Since at that time the stresses given in eqn. (25) did 
not accord to any acceptable physical distribution, it was not further 
pursued.  \\
  
    \n Further it is known that the equation of state $\rho + 3p = 0$, 
which means $E=0$, characterizes global 
    texture  [14-15]. That  is, the necessary condition  for spacetimes of 
   topological defects,  global  textures  and monopoles, is   $E = 0$. 
Like the uniform potential spacetime, it can also be shown that the 
global texture spacetime is dual to flat spacetime.  In the above 
eqns (30) and (31), replace $w_a w_b$ by the projection tensor $h_{ab} = 
g_{ab} - u_a u_b$. Then   
   non-static homogeneous solution of eqn.\ (30) is flat while that of the  
dual-flat equation (31) is the FRW 
   metric with $\rho +3p = 0$, which  
   determines  the scale factor $S(t) = \alpha t + \beta, $ and $\rho = 3
    (\alpha^2 +$ k) $/ (\alpha t + \beta)^2, $ k $= \pm 1, 0$. This is also 
   the unique non-static homogeneous solution. The general solutions of the 
   dual-flat equation are thus the massless global monopole (uniform 
   potential) spacetime in the static case and the global texture spacetime 
   in the non-static homogeneous case. Thus they are dual to flat 
   spacetime.  \\
     
    \n It  turns  out  that spacetimes  with $E=0$  
    can  be 
    generated [16]  by  considering  a  hypersurface   in   5-dimensional 
    Minkowski space defined, for example, by 
    
    \be
    t^2 - x^2_1 - x^2_2 - x^2_3 - x^2_4 = k^2 (t^2 - x^2_1 - x^2_2 - x^2_3)
    \ee

    \n which consequently leads to the metric
    
    \be
    ds^2 = k^2 dT^2 - T^2 [d \chi^2 + \sinh^2 \chi (d \theta^2 + \sin^2 \theta 
    d \varphi^2)].
    \ee

    \n Here $T^2 = t^2 - x^2_1 - x^2_2 - x^2_3 $ and $\rho = 3(1-k^2)/k^2 T^2$.                               The above construction  will  
    generate spacetimes of global monopole, cosmic strings (and their 
    homogeneous versions as well), and global texture-like type depending 
    upon  the  dimension  and character of the  hypersurface. Of 
    course, $E=0$   always; i.e.  zero   gravitational   mass 
    [16]. The  trace of the active part measures the gravitational  charge 
    density,   responsible  for  focussing  of  congruences   in   the 
    Raychaudhuri  equation  [17]. The topological  defects are thus 
   characterized by vanishing of focussing density (tracelessness of active 
   part). 
    
     \n Application of the duality transformation, apart from vacuum/flat 
    case  considered  here, has been considered for  fluid  spacetimes 
    [7]. The duality transformation could similarly be considered for    
    electrovac equations including the $\Lambda$-term. Here the analogue
    of the master equation (23) is
    
    \be
    H_{[ab]} = 0, E = \Lambda - \frac{Q^2}{2r^4}, ~ E^b_a + 
    {\tilde E}^b_a = (- \frac{Q^2}{r^4} + \Lambda) w_a w^b
    \ee
    
    \n which has the general solution $c^2 = a^{-2} = (1-2k-2M/r
    + Q^2/2r^2 - \Lambda r^2/3)$ and $\Lambda = 2k/r^2$. The
    analogue of eqn. (22) will have ${\tilde E} = - \Lambda - Q^2/2r^4$
    instead of $E$ in (35). Thus the duality transformation works in general
    for a charged particle in the de Sitter universe [18]. Similarly 
a spacetime 
dual to the NUT solution has been obtained [19]. In the case of the Kerr 
solution it turns out, in contrast to 
others, that the dual solution is not unique. The dual equation admits two 
distinct solutions which include the original Kerr solution [20].

 \section{Discussion}

 \n It is remarkable that the Maxwell-like duality transformation of the 
electric and magnetic parts of gravitational field which is a 
symmetry of the Einstein action leads to the vacuum field equation. In GR 
the equation of motion of free particles is implied by the field equation, 
and now we have the equation of motion of the field being implied by the 
duality symmetry of the action. Since the Riemann curvature contains the 
second derivative of the metric, hence the dynamics of the field that 
would permeate electromagnetic parts as well. Thus a duality relation between
electric and magnetic parts that keeps the Einstein action invariant should 
imply some specific equation between them which could be nothing other  
than the equation of motion of the field. This property though seems very 
natural and to some extent obvious has not been, as far as I know, 
noticed earlier. This is the primary avatar of duality. The another 
important point to note is that this duality implies the vacuum equation 
without the cosmological connstant. The equation with $\Lambda$ is 
characterized by equality of the left and right dual of the Riemann 
curvature. But it is not a symmetry of the action. \\

 \n The second avatar is the one that keeps the vacuum equation 
invariant. It means interchange of active and passive electric parts and 
we term this gravo-electric duality. Since the equation remains 
invariant, so would vacuum solutions. It however turns out that 
the Weyl tensor and scalar curvature change sign. Thus invariance of the 
action would require $G$ to change sign, implying gravity changing its 
sense. It has been argued that the sources for active and passive parts 
are respectively non-gravitational energy and gravitational field energy. 
It is well-known that they are of opposite sign. Since active and passive 
parts are interchanged under duality and hence so would be their sources 
which are of opposite signs. Thus the gravitational constant must change 
sign and with this all vacuum black hole solutions are self dual. \\

 \n In the third avatar of duality, we have constructed distinct solutions 
dual to the well-known black hole solutions by modifying the vacuum 
equation which no longer remains invariant under the gravo-electric 
duality. The modified equation would still admit the unique black hole 
solutions because the modification is effected in the equation which was 
free, implied by the other. The dual solutions to 
Reissner-Nordstr\"{o}m, NUT and Kerr black holes have been found 
[4,18,19,20]. It also turns out that the de Sitter is dual to the anti de 
Sitter.\\

 \n Let us consider the simplest and most instructive case of the 
Schwarzschild solution. As we have seen above,  ultimately the 
vacuum equation reduces to the Laplace equation and its first integral. 
The latter knocks off the constant of integration in the solution of the 
former which was free in 
the Newtonian theory to fix zero of the potential. The Einstein equation 
does not sustain this freedom and fixes zero at infinity implying 
asymptotic flatness of the Schwarzschild spacetime. The dual solution on 
the other hand does nothing else than restoring this constant and 
breaking asymptotic flatness in the most harmless manner. Of course the 
spacetime would no longer be empty. The stresses generated by the 
constant are precisely the same that required for representation of a 
global monopole charge on a Schwarzschild particle [12]. The dual 
solutions thus retain the basic physical features of the original 
vacuum solutions. With the exception of the Kerr solution, the dual 
solutions to all other black holes are also unique and could be 
interpreted as black holes with global monopole charge. \\

 \n In the dual solution if we set the Schwarzschild mass to zero, the 
resulting spacetime would describe the field of uniform gravitational 
potential or of zero mass global monopole charge. The spacetime is 
obviously non-flat. Thus constant relativistic potential has non-trivial 
dynamics. This is because in GR potential does two things, one as the 
Newtonian potential as it appears in $g_{00}$ and other the relativistic 
effect of "curving" space in $g_{11}$. The former as expected can be 
transformed away while the latter cannot be even when potential is 
constant. Thus uniform potential produces non-zero curvature which could 
be envisioned as an example of "minimum" curvature. \\
 
 \n Since duality breaks asymptotic flatness without significantly 
altering the physical character of the field, it could be the most 
appropriate way to incorporate Mach's principle. To let the rest of the 
Universe be non-empty, it is of primary importance to break asymptotic 
flatness of spacetimes representing isolated bodies. At the same time 
the basic character of the field must not change. This is precisely what the 
dual solution does [18]. Consider the solar system sitting in the uniform 
potential of our galaxy. The constant in the dual to Schwarzschild solution 
would then be determined by the uniform galactic potential. In view of 
this, it can be argued that it is 
the dual solution that would perhaps describe the solar system more 
appropriately than the Schwarzschild [8]. The observational support to 
the Schwarzschild solution will also extend to the dual solution as
well [21]. Thus the
dual solution would be Machian at the very primary level. \\

 \n The modified vacuum equation (22) works all right for isolated 
sources and its dual gives the dual solutions. Can this be a true 
characterization of vacuum? The equation has $\tilde E = 0$ which means 
vanishing of energy density. This plus absence of energy flux should lead 
to vacuum. For non-isolated cases, it turns out that eqn.(22) admits the 
well-known vacuum solutions describing gravitational wave, the Weyl and 
Levi-Civita metrics. The dual equation (23) too admits the same vacuum 
solution. They are all self-dual [22]. \\   

 \n Finally we would like to say that application of the gravo-electric 
duality transformation is not confined to GR alone. It can also be 
applied to construct solutions dual to stringy black holes with dilaton field 
[23], as well as in the 2+1 gravity [24]. Work on studying its 
application in other theories is in progress.\\

      \n {\bf Acknowledgement  :}  I wish to thank Jose  Senovilla, 
 and  LK Patel and Ramesh  Tikekar for useful discussions. Above all
it is a matter of great pleasure to dedicate this work to George, who 
has brought forth the study of Weyl electromagnetics and its application 
in cosmology, on his youthful 60.\\


\newpage
    \begin{thebibliography}{99}
    
\bibitem{} L. Bel (1958) C.R.Acad.Sci. {\bf246},3105.    
    
\bibitem{} M.A.G. Bonilla and J.M.M. Senovilla (1997) Gen.Rel.Grav. 
{\bf 29},91.
    
\bibitem{} N.  Dadhich in {\it Black holes, Gravitational Radiation and 
the Universe}, eds. B.R. Iyer and B. Bhawal (Kluwer, 1999), p.171.

\bibitem{} ----------- (1999) Mod.Phys.Lett. {\bf A14},337.

\bibitem{} ----------- (1999) Mod.Phys.Lett. {\bf A14},759.

\bibitem{} --------  in {\it Gravitation and Relativity in general}, eds. 
A. Molina, J. Martin, E. Ruiz, and F. Atrio (Proceedings of ERE-98, 
Salamanca) to be poublished by World Scientific.   

\bibitem{} N.   Dadhich,   L.K.   Patel  and   R.   Tikekar   (1998) 
   Class. Quantum Grav. {\bf 15}, L27. 
    
\bibitem{} N. Dadhich (1997) On the Schwarzschild field, gr-qc/9704068.

\bibitem{} S. Chandrasekhar (1983) {\it The Mathematical Theory of Black 
Holes}, (Oxford).

\bibitem{} D. Lynden-Bell and M. Nouri-Zonoz (1998) Rev.Mod.Phys. {\bf 
70},427.

\bibitem{} C.W. Misner (1963) J.Math.Phys. {\bf4},924.

\bibitem{} M. Barriola and A. Vilenkin (1989) Phys.Rev.Lett. {\bf 63},341.

\bibitem{} N. Dadhich (1970) Ph.D. Thesis, Poona University, unpublished.   

\bibitem{} R.L. Davis (1987) Phys.Rev. {\bf D35},3705.

\bibitem{} D. Notzold (1991) Phys.Rev. {\bf D43}, R961.

\bibitem{} N. Dadhich and K. Narayan (1998) Gen. Rel. Grav. {\bf 30}, L1133.

\bibitem{} A.K. Raychaudhuri (1955) Phys.Rev. {\bf 90}, 1123.

\bibitem{} N. Dadhich, Dual spacetimes, Mach's principle and 
topological defects, gr-qc/9902066.

\bibitem{} M. Nouri-Zonoz, N. Dadhich and D. Lynden-Bell (1999) 
Class.Quant.Grav. {\bf16},1021.
    
\bibitem{} N. Dadhich and L. K. Patel (1999) Gravo-electric dual of 
the Kerr solution, submitted.
   
\bibitem{} N. Dadhich, K. Narayan and U.A. Yajnik (1998) Pramana {\bf 50}, 307.
   
\bibitem{} L.K. Patel and N. Dadhich, On dual of cosmological vacuum 
solutions, to be submitted. 

\bibitem{} S. Bose and N. Dadhich, On the gravo-electric dual of the 
static charged black hole solutions in string theory, to be submitted.
     
\bibitem{} S. Bose, N. Dadhich and S. Kar, On gravo-electric duality in 2+1 
gravity, to be submitted.

   
      \end{thebibliography}
    \end{document}